\documentclass[aps,prl,twocolumn,superscriptaddress,showpacs]{revtex4}

\usepackage{graphicx}

\begin{document}

\title{Electronic stopping power in insulators from first principles}

\author{J. M. Pruneda}
\affiliation{Instituto de Ciencia de Materiales de Barcelona 
(ICMAB-CSIC) Campus de Bellaterra, 08193 Barcelona, Spain}
\affiliation{Department of Physics, University of California,
Berkeley, CA 94720}
\author{D. S\'anchez-Portal}
\affiliation{Donostia International Physics Center, Paseo Manuel de 
Lardizabal 4, 20018 San Sebastian, Spain}
\affiliation{Centro Mixto CSIC-UPV/EHU, Facultad de Qu\'{\i}mica, 
Apartado 1072, 20080 San Sebastian, Spain}
\author{A. Arnau}
\author{J. I. Juaristi}
\affiliation{Donostia International Physics Center, Paseo Manuel de 
Lardizabal 4, 20018 San Sebastian, Spain}
\affiliation{Centro Mixto CSIC-UPV/EHU, Facultad de Qu\'{\i}mica, 
Apartado 1072, 20080 San Sebastian, Spain}
\affiliation{Departamento de F\'{\i}sica de Materiales, Facultad de 
Qu\'{\i}mica, Apartado 1072, 20080 San Sebastian, Spain}
\author{Emilio Artacho}
\affiliation{Department of Earth Sciences, Downing Street,
University of Cambridge, Cambridge CB2 3EQ, United Kingdom}
\affiliation{Donostia International Physics Center, Paseo Manuel de 
Lardizabal 4, 20018 San Sebastian, Spain}

\date{\today}

\begin{abstract}
  Using time-dependent density-functional theory we calculate from first
principles the rate of energy transfer from a moving proton or antiproton 
to the electrons of an insulating material, LiF.
  The behavior of the electronic stopping power versus projectile velocity
displays an effective threshold velocity of $\sim 0.2$ a.u. for the proton,
consistent with recent experimental observations, and also for the 
antiproton.
  The calculated proton/antiproton stopping-power ratio is $\sim 2.4$ at 
velocities slightly above the threshold ($v\sim 0.4$ a.u.), as compared 
to the experimental value of 2.1.
  The projectile energy loss mechanism is observed to be stationary and 
extremely local.
\end{abstract}

\pacs{61.85.+p; 61.80.Jh; 61.82.Ms; 34.50.Bw; 71.15.Pd}

\maketitle


  The interaction of charged particles with matter has been a subject of 
extensive research since the discovery of subatomic particles~\cite{Review90}.
  Ions moving through solids gradually transmit their kinetic energy 
to electronic excitations of the host and deposit it along their path. 
  The maximum of this deposited energy is the so-called Bragg peak and 
it occurs shortly before the particle stops.
  Hence the importance of studying the electronic energy loss of slow 
ions (with velocities below the Bohr velocity) travelling through solids.
  For metals, the understanding of this problem has been steadily progressing 
over the years~\cite{Review90}.
  For insulators, however, experimental results remain unexplained even 
for simple systems.
  This is particularly true at low velocities (the threshold effect) 
\cite{Eder97,Bauer05}, where the contribution from nuclear 
collisions conceals the electronic stopping \cite{lattice-scattering}. 

  A detailed quantitative knowledge of these processes is required to 
understand the damage produced in materials when exposed to radiation. 
  For ceramic materials devised for the encapsulation of 
nuclear waste \cite{Farnan07} the prediction of durability over extremely 
long times is crucial. 
  Radiation-damage simulations performed to date \cite{Kostya06,Grimes04} rely 
on empirical force fields obtained from fits to low-energy properties. 
  The actual interatomic forces could be enormously altered, however, by the 
local electron heating produced by the electronic stopping.


  In the semiclassical formalism, the electronic energy loss rate is given 
by the response of the system to the external potential,
$\frac{dE}{dt}=-Z{\bf v}\cdot{\bf E}^{ind}$,
where $Z$ and ${\bf v}$ are the charge and velocity of the projectile, 
and ${\bf E}^{ind}$ is the induced electric field in the target material.  
  Hence, to first order, linear response theory can be used to give the 
stopping power (SP) --the energy lost per unit path length-- in terms of the 
dynamical dielectric function of the material. 
  For metals this formalism shows that the SP is linear with small velocities,
$\frac{dE}{dx}\sim (Ze)^2v$~\cite{Ritchie59,Kitagawa73}, reflecting that no 
minimum energy is required to excite electron-hole pairs.
  It must be remembered, however, that the dielectric approach is not valid at 
low ion velocities, and non-linear effects cannot be neglected~\cite{Review90}.


  A different behavior is expected in wide band-gap insulators, given the
finite energy required to excite electrons.
  For protons moving through noble gases~\cite{Deumens00}, which also display 
a large minimum excitation energy, a velocity threshold has been observed 
experimentally \cite{Golser91}, and explained in terms of the quantization of 
energy levels and charge exchanges. 
  However, for solid insulators the situation is unclear. 
  No threshold effects were originally observed in Al$_2$O$_3$, SiO$_2$, 
or LiF, and the linear dependence $dE/dx \propto v$ was observed down to 
velocities of about 0.3 a.u.~\cite{Eder97,Juaristi00,Moller}.

  For protons under grazing incidence in LiF(001), and below 
$\sim 0.2$ a.u. a threshold behavior was reported~\cite{Auth98}.
  Under these conditions the proton does not penetrate the solid, and 
charge exchange is identified as the dominant mechanism for electronic 
stopping, with local electron capture from F$^-$ ions, giving rise to 
H$^0$ and H$^-$.  
  More recent experiments on thin LiF films show an apparent velocity 
threshold near 0.1 a.u.~\cite{Bauer05}.
  The different experimental setup (transmission) suggests a different
physical stopping mechanism, based on electron-hole pair excitations.
  A threshold behavior is expected, and has been qualitatively predicted
from linear-response and from perturbation theory calculations. 
  These approximations, however, grossly underestimate the SP
at low velocities, thus exaggerating the threshold~\cite{Inhaki}.
  A theoretical description beyond these approximations is needed.


  In this letter we present a first-principles approach to the 
non-perturbative study of realistic solid-ion interactions. 
  We follow the explicit time evolution of the electronic states of the host 
crystal as an external particle propagates through the system, by means
of time-dependent density-functional theory (TD-DFT)~\cite{td-dft}.
  We use here this scheme to understand the SP threshold effect and stopping 
mechanisms in LiF, a well-studied characteristic insulator, finding 
reasonable agreements with measured data.
  Most importantly, however, this study sets the scene for a promising
line of theoretical simulations, assesses its possibilities and offers 
new insights into the stopping process.


  The energy transmitted to the electrons from a constant-velocity moving
ion is monitored.
  Total energy is thus not conserved since the energy loss of the projectile
will be neglected (its large mass ensures a negligibly small decline in its
velocity on the time scales of the simulations).
  As the center of the charged particle, ${\bf r_c}$, moves with constant 
velocity, ${\bf r_c}={\bf r_0}+{\bf v}\cdot t$, the time-dependent 
Kohn-Sham (KS) equation~\cite{td-dft} defines the dynamics of effective 
single-particle states (and thereby the electronic density and energy) 
under the external potential generated by the projectile and the crystal 
of Li and F nuclei.

  The calculations were done using the {\sc Siesta} {\it ab initio} 
method~\cite{Ordejon96,siesta}, in its time-evolving TD-DFT 
implementation~\cite{DSP02}, using the instantaneous local density 
approximation (LDA) to exchange and correlation~\cite{LDA}.
  The 1$s$ core electron pair of F was replaced by a norm-conserving 
pseudopotential~\cite{tm} in the fully nonlocal form.
  The 1$s$ electrons of Li were explicitly included in the calculation as 
pseudized valence electrons.
  A double-$\zeta$ polarized basis was considered for the valence 
electrons (single-$\zeta$ for the 1$s$ of Li), with an energy shift of
100 meV \cite{siesta}. 
  The grid cutoff \cite{siesta} for integration was 118 Ry.
  The lattice parameter obtained for bulk LiF is 3.98~\AA, slightly smaller 
than the experimental value 4.03~\AA, as expected for LDA. 
  The projectile (proton, $p$, or antiproton, $\bar{p}$) was described by the bare 
$\pm\frac{1}{r}$ potential 
\footnote{The electrostatic energy divergence
due to the charge of the projectile in periodic boundary conditions is avoided
by using $\frac{e^{-r/\rho}}{r}$ instead of $\frac{1}{r}$ and converging
the SP with respect to $\rho$.
  This is equivalent to a compensating homogeneous background.
  The finite-cutoff filtering of the $\frac{1}{r}$ cusp customary in plane-wave
calculations \cite{Vandewalle89} has been replaced here by a Gaussian charge
smoothening of a width dictated by the grid cutoff used in {\sc siesta}
\cite{siesta}.  }.
  It was not pseudized in order to treat $p$ and $\bar{p}$
on the same footing.

\begin{figure}[t]
\includegraphics[width=0.40\textwidth]{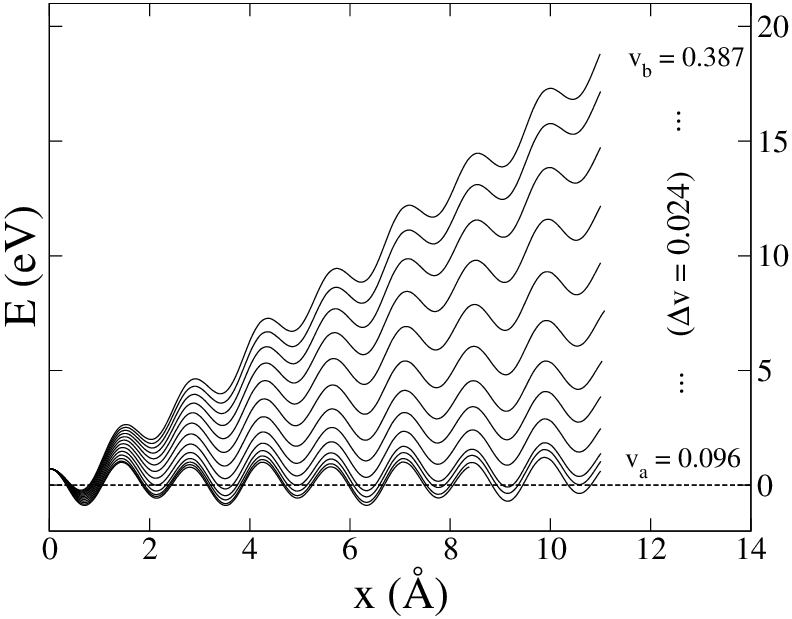}
\caption{\label{Evst} Total electronic energy as a function of displacement
for several proton velocities. }
\end{figure}

  Periodic boundary conditions were used throughout.
  The supercell size was chosen so as to minimize the spurious effects of
the repetition while keeping manageable computational demands.
  After convergence tests, a 4x4x4 supercell of 128 atoms was selected,
such that a particle moves $\sim$11\AA\ down a [110] channel before 
re-entering the box.  
  The $\Gamma$-point approximation was used for integrations in $k$.
  The projectile was initially put in the centre of a crystal cage
and the time-independent DFT solution was obtained
to define the initial state for the subsequent evolution.
  It was then moved with constant velocity along a [110] channel. 
  The time-dependent KS equation was then solved numerically by 
discretizing time and applying the Crank-Nicholson algorithm as described 
in ref.~\cite{DSP02}.  
  Using a time step of 1 attosecond, the wavefunctions were then
propagated for several femtoseconds. 
  The electronic total energy was recorded as a function of time and the 
SP, $\frac{dE}{dx}$, was extracted.

  Fig.~\ref{Evst} shows the electronic total energy of the system as a 
function of displacement for different velocities of the projectile. 
\begin{figure}[b!]
\includegraphics[width=0.40\textwidth]{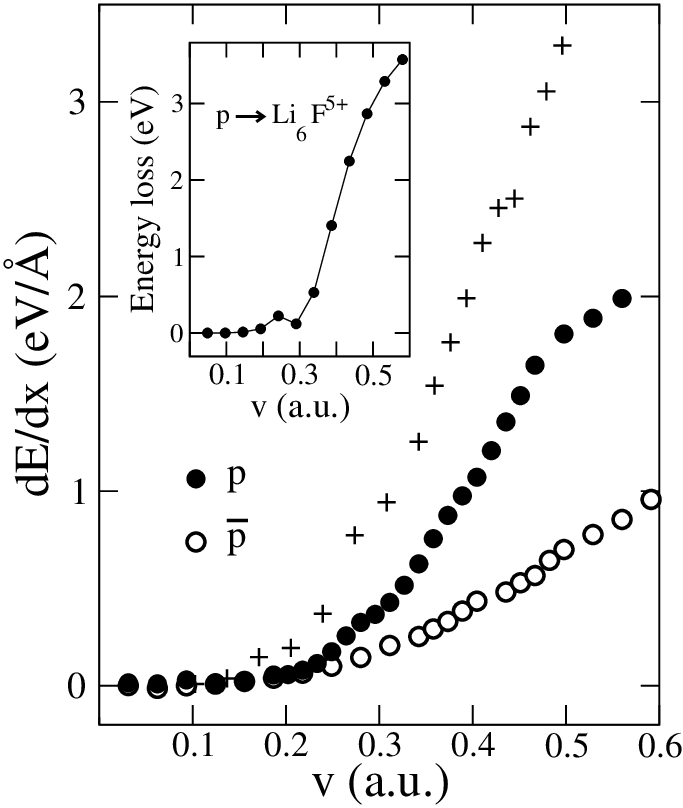}
\caption{\label{stopping} Calculated electronic stopping power 
$\frac{dE}{dx}$ as a function of particle velocity $v$ for $p$
(empty circles) and $\bar{p}$ (full circles). 
  Crosses indicate calculations for $p$ including additional basis 
orbitals along the projectile's path (see text).
  The inset shows the excess energy acquired by a octahedral Li$_6$F$^{5+}$
cluster on the passage of a $p$ on an analogous trajectory to the solid's.}
\end{figure}
  At low velocities the adiabatic behavior is recovered, with no
net energy transfer, just the expected oscillation of the total energy
with the position of the projectile in the crystal.
  At higher velocities the oscillations are superimposed to an underlying 
energy increase with time. 
  After a remarkably short transient period (around 0.3 fs), the energy 
increase stabilizes to an essentially stationary regime, in 
which the energy difference between consecutive equivalent lattice positions 
of the projectile remains constant.
  The SP is then extracted from the average slope of this 
stationary regime.
  Stopping power results for $p$ and $\bar{p}$ are presented in 
Fig.~\ref{stopping}.  
  A threshold effect at low velocities is apparent in the figure, in 
agreement with recent experiments~\cite{Bauer05}, unlike the linear 
behavior observed for metals~\cite{Review90}, but much smaller than
that predicted for insulators by linear response theory~\cite{Inhaki}.
  The values obtained in this work (around 0.2 a.u.) are consistent 
with experiments~\cite{Bauer05,Moller}.
  The difference between $p$ and $\bar{p}$ is also apparent,
in contrast with the invariance under charge-sign change expected from 
perturbative treatments. 
  Experiments show that the energy loss for protons is approximately twice
as high as for antiprotons~\cite{Moller}, consequence of the different kind 
of screening (one attracts electrons while the other attracts holes).
  In the neighborhood of $v=0.4$ a.u. (slightly above the threshold) 
we obtain a SP ratio of $\sim 2.4$, that compares well with the experimental 
value of 2.1~\cite{Moller}.

\begin{figure}[b!]
\includegraphics[width=0.35\textwidth]{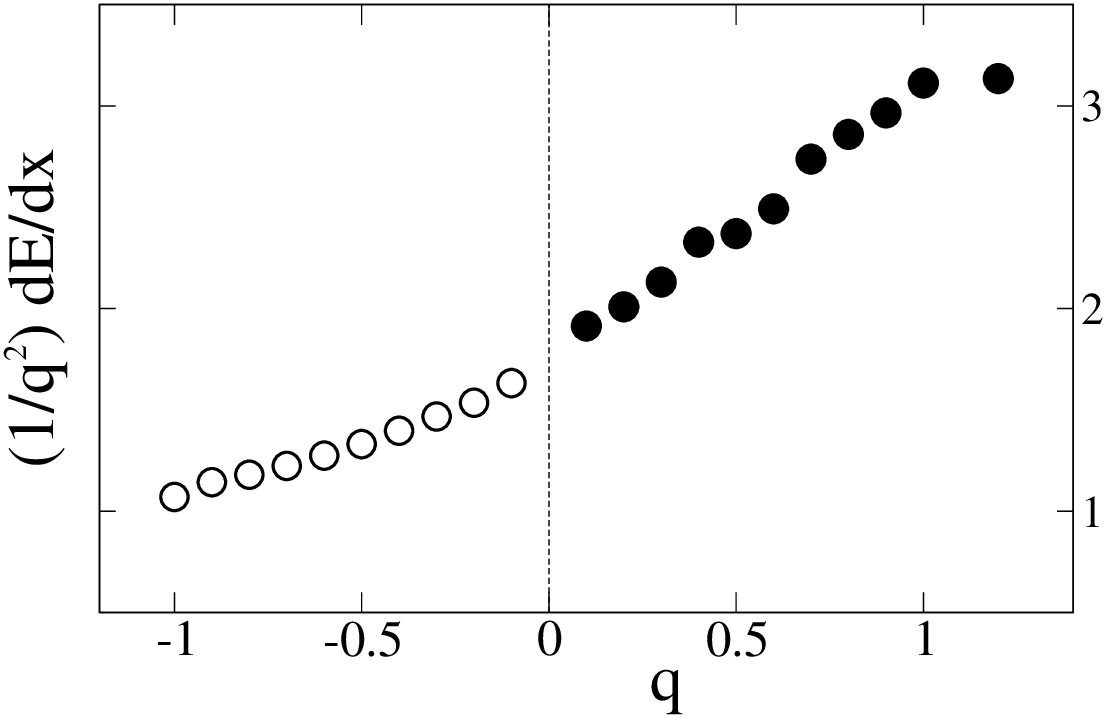}%
\caption{\label{charge} Dependence of $\frac{1}{q^2}\frac{dE}{dx}$ with the 
charge $q$ of the external projectile, for $v=0.46$ a.u. 
  The stopping power is in eV/\AA, and $q$ in electron charge units.}
\end{figure}

  Since a $q^2$ charge dependence is expected from linear response, the SP
over $q^2$ is plotted in Fig.~\ref{charge} for both projectiles.
  The smooth behavior at the origin indicates that there are no substantial
biases in the way the stopping mechanism is described for $p$
and $\bar{p}$.
  The slope near the origin corresponds to the $q^3$ dependence, while
the bending corresponds to $q^4$ and higher terms.
  The displayed behavior, and particularly the sign of both terms,
is similar to that observed for metals~\cite{Echenique91}.

  There is a discrepancy between measured and computed values for
the SP for velocities above the threshold.
  The experiments show four times higher SP~\cite{Bauer05}. 
  A factor of two is accounted for by the known relation between channelling 
conditions and the average over random trajectories~\cite{Grande91,Dorado93}.
  Technical reasons account for a part of the remaining discrepancy.
  The inclusion of basis orbitals along the projectile's path (a 
$sp$ single-$\zeta$ set every 0.5 \AA) increases the SP
from 1.6 eV/\AA\ to 2.8 eV/\AA\ for $v=0.46$ a.u.
\footnote{
  The basis set used throughout this work has been the one defined by
atomic orbitals of target atoms, so that no explicit bias was introduced 
in the comparison of $p$ and $\bar{p}$.
  There is a finite-basis saturation effect at high velocities, but that 
is observed at higher velocities than the ones studied here.
  Although the absolute value of the SP is not completely 
converged by our basis set, the convergence for both the relation 
proton/antiproton and the respective thresholds is satisfactory.}
as shown in Fig.~\ref{stopping}. 
  At a more fundamental level, discrepancies are expected
from the errors in electronic spectra around the band gap given by 
instantaneous LDA in TD-DFT (even if it includes a RPA correction on 
KS eigenvalues)~\cite{Gross03}.

\begin{figure}[t]
\includegraphics[width=0.35\textwidth]{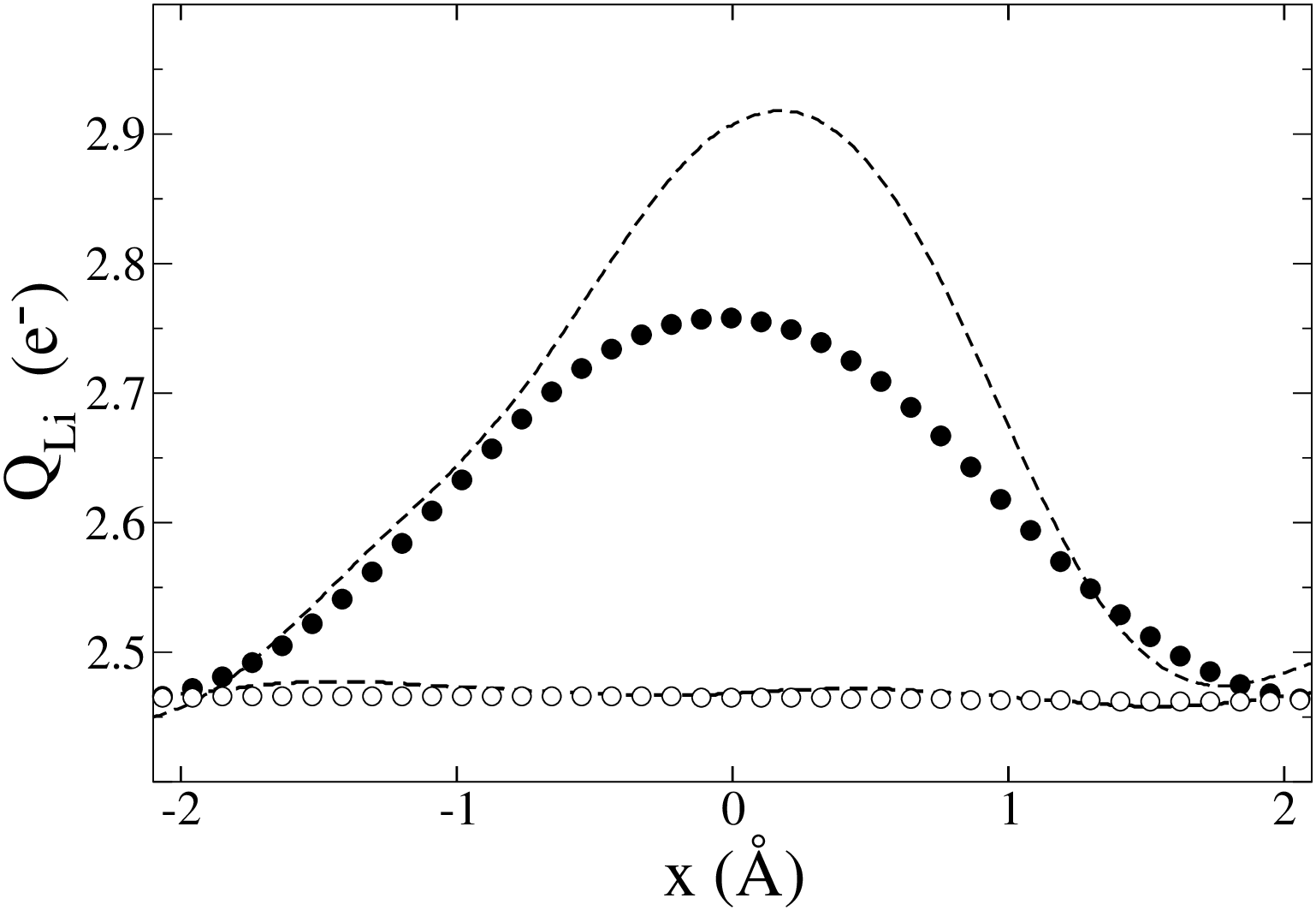}
\caption{\label{local} Electron population on Li atoms close to the 
proton trajectory versus position of the projectile along it. 
Adiabatic (dynamic) charges are shown by circles (dashed lines).
The position of the maximum in the adiabatic curve occurs when the
projectile is closest to the atom ($\sim$1.2\AA).  The almost 
constant line corresponds to a second closest Li atom, 2.3\AA\ 
away from the trajectory.}
\end{figure}

  We now analyze the locality of the energy absorption in terms of Mulliken
charges~\cite{Mulliken}.
  Although they are known to be unreliable in their explicit dependence on 
the basis set, their changes for fixed basis set are much more indicative. 
  By comparing the atomic charges obtained dynamically and 
adiabatically (a static self-consistent calculation with the external 
potential at the same instantaneous position), insights are obtained into 
the stopping process.
  Fig.~\ref{local} shows this comparison for the proton case, where
the following is observed:
  ($i$) The screening of the proton is enhanced in the dynamic case (our
basis set describes this screening by a transfer of charge from F to Li),
due to the increased polarizability of the host at the frequencies explored
by the moving ion.
  ($ii$) A delay is apparent in the dynamical screening, which is the main 
responsible of the dissipation.
  ($iii$) The electronic excitation process is extremely local: the changes
observed in atoms closest to the trajectory are much larger than
any other, and the dynamical screening effect is only noticeable when the
projectile is very close to the ion.

  The locality of the energy transfer is confirmed when calculating the
energy absorbed by a small cluster of LiF (Li$_6$F$^{5+}$). 
  The inset of Fig.~\ref{stopping} shows a striking similarity in the
overall $v$-dependence of the energy absorbed by the cluster and the
SP in the solid.
  If we took the effective path length in the cluster as 1.4 \AA\ (the
number of valence electrons in the cluster corresponds to one formula unit), 
both SPs would be indeed very similar
\footnote{TD-DFT is known to provide better results for small systems~\cite{Gross03}, we thus expect the spectra of our cluster to be more accurate than that of bulk.}.
  In the solid, the energy accumulated along the path would then diffuse away 
at longer time scales (corresponding to the effective band width), 
defining a wake.
  A tentative definition of local energy
\footnote{The energy associated to a basis function $|\mu \rangle$ is defined
as $E_{\mu} = \sum_{\nu} \rho_{\mu\nu} H_{\nu\mu}$, where $\rho_{\mu,\nu}$ and
$H_{\mu\nu}$ are the density and Hamiltonian matrices, respectively.
  Note that it does not include double-counting terms.}
shows well differentiated time scales for the 
excitation by the projectile, on one hand, and the ensuing out-diffusion, 
on the other (not shown).
  The short-ranged initial excitation can be rationalized
in terms of the electronic localization length scale relevant to dielectric
response~\cite{Souza00}, which is expected to be very short for LiF.
  Similar locality in the energy loss mechanism has been recently found in 
a confined two-dimensional electron gas, however~\cite{Borisov06}.

  A characterization of the charge state of the 
projectile~\cite{Narmann90,Prokscha07} has not been attempted. 
  The present calculations allow for the establishment of charge states
of any kind within the constraints that the electronic charge and momentum 
are conserved.
  It is clear that the mid-gap state travels with the proton, and that
it becomes partly populated.
  It is not clear, however, that the charge associated to that state 
represents a meaningful definition of the charge state of the projectile,
since whether the screening charge builds up as it passes or travels with it
is not determined by the population of that state.
  This effect will be explored in further works.


  In conclusion, we have presented a general approach for the non-perturbative
first-principles study of the electronic stopping power in solids.
  New insights into the electronic SP in insulators have been provided for 
protons and antiprotons in LiF.

\begin{acknowledgments}
We thank P. M. Echenique, E. Salje and I. Nagy for interesting discussions. 
EA acknowledges the hospitality of the Donostia International Physics Centre.
This work has been funded by the EC's Marie Curie program, UK's BNFL and NERC, 
the Spanish MEC, the UPV/EHU and the Basque 
Government through the Etortek program.
\end{acknowledgments}

\bibliography{stopping}

\end{document}